\documentclass[12pt]{article}
\usepackage{amsmath}
\usepackage{graphicx}
\usepackage{geometry}
\usepackage[applemac]{inputenc}

\usepackage{natbib}

\ifx\pdfoutput\@undefined\usepackage[usenames,dvips]{color}
\else\usepackage[usenames,dvipsnames]{color}
\IfFileExists{pdfcolmk.sty}{\usepackage{pdfcolmk}}{} 
\fi
\usepackage[plainpages=false,pdfpagelabels,pagebackref=false,naturalnames=true,hyperindex=true,pdftitle={***},pdfauthor={Carlos Gershenson}]{hyperref}
\hypersetup{colorlinks=true,
urlcolor=Cerulean,linkcolor=BrickRed,citecolor=RoyalBlue,
  pdfpagemode=None,
  pdfstartview=FitH}
\usepackage[all]{hypcap}

\begin{document}

\title{Adaptive Cities:\\ A Cybernetic Perspective on Urban Systems}

\author{Carlos Gershenson$^{1,2,3,4,5}$, Paolo Santi$^{3,6}$, and Carlo Ratti$^{3}$\\
$^{1}$Instituto de Investigaciones en Matem\'aticas Aplicadas y en Sistemas,\\
 Universidad Nacional Aut\'onoma de M\'exico.\\
$^{2}$ Centro de Ciencias de la Complejidad,\\ Universidad Nacional Aut\'onoma de M\'exico.\\
$^{3}$ SENSEable City Lab, Massachusetts Institute of Technology, USA. \\
$^{4}$  MoBS Lab, Network Science Institute, Northeastern University, USA.\\
$^{5}$ ITMO University, Russian Federation.\\
$^{6}$ Istituto di Informatica e Telematica, CNR, Italy.
}

\maketitle

\begin{abstract}
Cities are changing constantly. All urban systems face different conditions from day to day. Even when averaged regularities can be found, urban systems will be more efficient if they can adapt to changes at the same speeds at which these occur. Technology can assist humans in achieving this adaptation. Inspired by cybernetics, we propose a description of cities as adaptive systems. We identify three main components: information, algorithms, and agents, which we illustrate with current and future examples. The implications of adaptive cities are manifold, with direct impacts on mobility, sustainability, resilience, governance, and society. Still, the potential of adaptive cities will not depend so much on technology as on how we use it.
\end{abstract}


\section{Introduction}

Cities have become central to our species, with an increasing majority of people living in them~\citep{Cohen:2003,Cities:2010} and producing most of the wealth of our globalized society~\citep{sassen2011cities,dobbs2011urban}. They serve as magnets for migration as they offer several advantages and opportunities over rural areas~\citep{Glaeser:2011,Bettencourt:2007,Bettencourt:2010}. Densification of population is desirable for a sustainable urban development. However, a high population density also generates several problems which we must face, better sooner than later. We can identify urban problems related to mobility~\citep{Gakenheimer1999671}, pollution~\citep{bulkeley2003cities}, sanitation~\citep{jacobi2010citizens}, segregation~\citep{musterd2013urban}, marginalization~\citep{adler1975como}, and crime~\citep{glaeser1996there}, just to name a few. 

Even when we are increasingly dependent on urban systems, they are becoming unmanageable with traditional techniques. This is because of the inherent complexity of cities. The term complexity comes from the Latin \emph{plexus} which means intertwined. A complex system is such that its elements are difficult to separate. As elements are interdependent, their future depends not only on initial and boundary conditions, but on the \emph{interactions} which take place in time and space, generating novel information~\citep{Gershenson:2011e}. This information generated by interactions limits predictability. Since traditional techniques (such as optimization) rely on predictability, they cannot cope with the increasing complexity of our urban systems. 

Complexity is increasing because interactions and interdependencies are increasing. A more connected system can have advantages, as information, energy, and matter can spreads faster through it, it can respond faster to changes~\citep{Khanna2016}. However, an increased connectivity also has its drawbacks: having many components affecting each other can potentially increase the fragility of a system~\citep{Taleb2012,Helbing2013Globally-networ}.

Given the complex nature of urban systems, they change constantly~\cite{Batty1971Modelling-Citie}, and thus problems change as well, \emph{i.e.} they are non-stationary~\citep{GershensonDCSOS}. This implies that trying to find optimized solutions will be inefficient, as the optimal solution changes with the problem. If traditional techniques cannot cope with the complexity and dynamics of urban systems, how can we regulate them? \emph{Adaptation} is required to let urban systems to change their behavior according to their current situation~\citep{Gershenson2013Facing-Complexi,Rauws22072016}. We have plenty of examples of adaptation in living systems, which can serve as an inspiration for urban solutions~\citep{Alexander:2003,Gershenson:2013}. 

\section{Cybernetics}

The study of adaptivity in systems began decades ago within cybernetics \citep{Wiener1948,Ashby1956,pask1961approach}. The relevance of cybernetics lies in the fact that it was the first scientific attempt to study phenomena independently from their substrate, \emph{i.e.}, focussing more on the function of systems than on their composition~\citep{Gershenson2013The-Past-Presen}. This allowed the cross-fertilization of different scientific fields, \emph{e.g.} electrical engineering and neuroscience, where similar functions are required by systems composed by different components.

\begin{figure}[htbp]
\begin{center}
\includegraphics[width=0.35\textwidth]{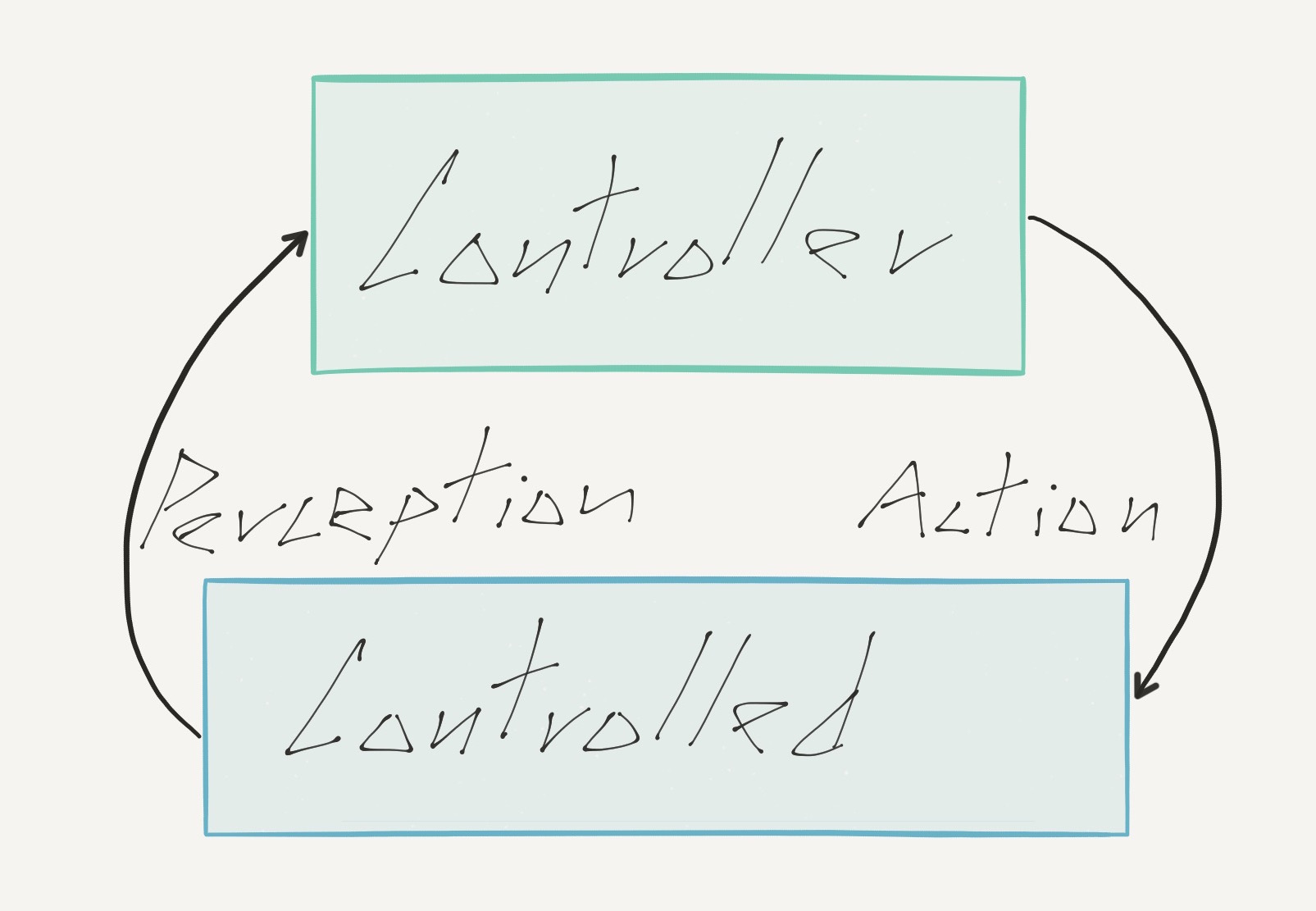}
\caption{An abstract cybernetic control loop.}
\label{fig:loop}
\end{center}
\end{figure}

One of the most used concepts from cybernetics is that of the control loop~\citep{HeylighenJoslyn2001}. As illustrated in Figure~\ref{fig:loop}, a controller perceives inputs from the controlled and acts with its outputs on the controlled. The controlled has its own dynamics, i.e. its variables are changing. That is why the controller must perceive, to detect the changes, make decisions, and take actions to keep the variables controller within a desired state. Note that control loops can take place at multiple scales: subunits, units, modules, systems, or metasystems.

\begin{figure}[htbp]
\begin{center}
\includegraphics[width=0.35\textwidth]{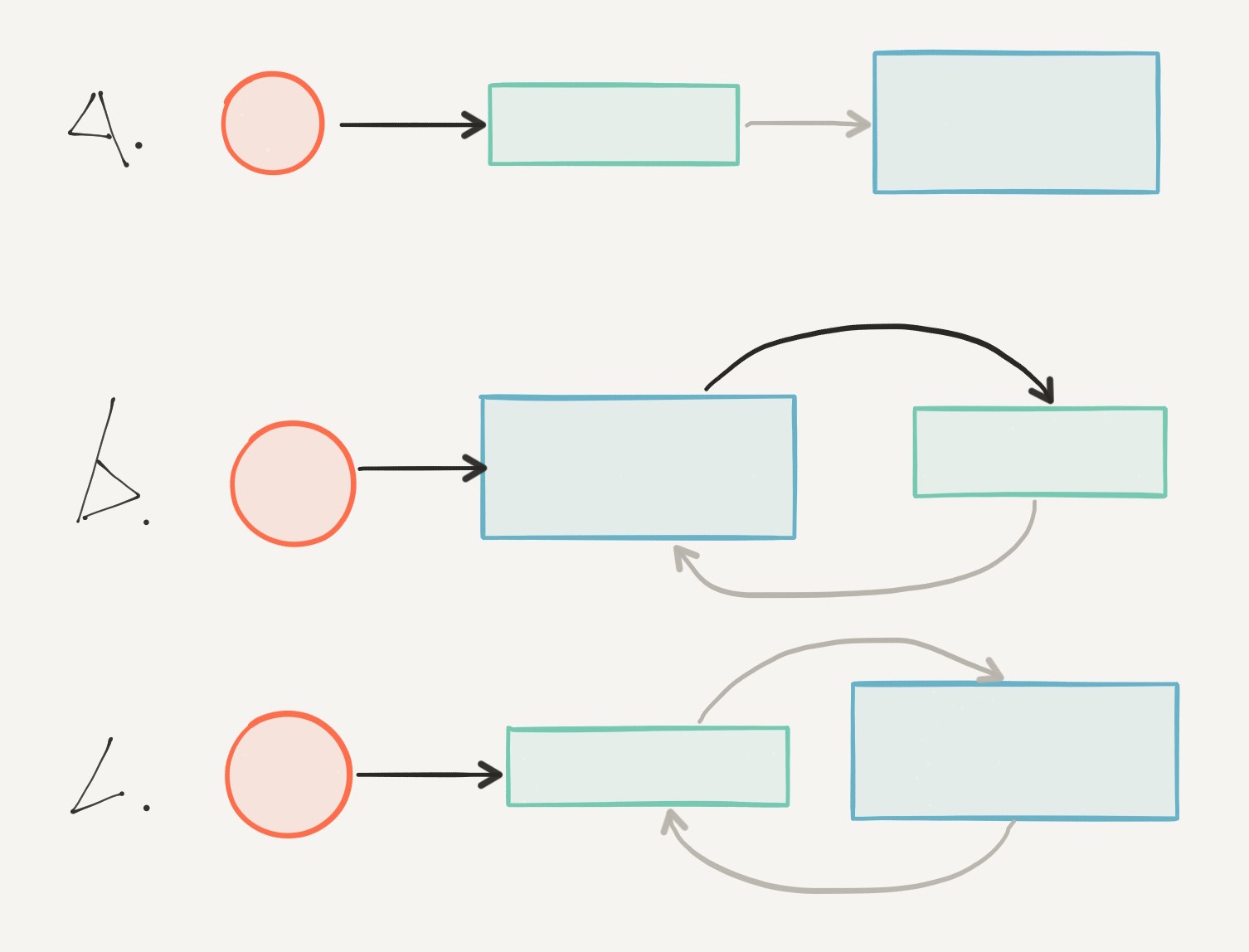}
\caption{Perturbations (circles) on systems (larger rectangles) can be controlled by a. buffering, b. feedback, or c. feedforward mechanisms (smaller rectangles).}
\label{fig:mech}
\end{center}
\end{figure}

In general, a controller will try to steer a system (controlled) towards a desired state, configuration, or behavior. Perturbations (internal or external) might deviate the system, so the controller should compensate those perturbations. This can be achieved by buffering, feedback or feedforward mechanisms. As shown in Figure~\ref{fig:mech}, these mechanisms can be used to counteract the effect of perturbations on the controlled systems. Buffering is \emph{passive}. It basically diminishes or nullifies the effect of perturbations. For example, insulation reduces the effect of temperature differences. Feedback mechanisms act after the system has been perturbed, trying to return the variables of the system to their desired state. For example, a thermostat can detect that temperature is lower than desired and switch on the heating until the desired temperature is reached. Feedforward mechanisms act before the perturbation manages to affect the system to prevent their effect. For example, if a smart thermostat knows that the temperature might decrease at night, it might switch on the heating before the temperature decreases, so it never leaves its desired state. Feedback and feedforward mechanisms are \emph{active}. A system can adapt to perturbations using these mechanisms.

We can define adaptation as ``a change in an agent or system as a response to a state of its environment that will help the agent or system to fulfill its goals"~\citep{GershensonDCSOS}. Adaptation can be achieved by reaction (feedback mechanisms) or anticipation (feedforward mechanisms). It would be desirable to predict all possible perturbations to a system, so that they could be handled before they can affect the variables of a controlled system using anticipation. However, since predictability is limited due to complexity~\citep{Morin2006}, we can always expect unexpected perturbations. For all the unpredictable perturbations, it is necessary to react \emph{a posteriori} using feedback mechanisms. 

To prevent the effect of perturbations, buffering can increase the robustness of systems. A system is robust if it continues to function in the face of perturbations~\citep{Wagner2005}. Robustness is desirable to minimize perturbations, but since change is unavoidable, adaptation (active control) is required. On the one hand, because all perturbations cannot be predicted (so as to build a perfect buffer). On the other hand, because too much robustness can limit adaptation~\citep{GershensonEtAl2006}. Moreover, adaptation can increase robustness, as adaptive change is made precisely to preserve the function of a system.

These and other cybernetic concepts have permeated into all disciplines. 
For example, Stafford Beer used cybernetic concepts to achieve adaptive organizations~\citep{Beer1966,GershensonSOBs}. This was applied at a national scale in Chile in the early 1970s with the Cybersyn project~\citep{Medina2011}, which served as a ``nervous system" for the country. Unfortunately, the system was dismantled in 1973 by the dictatorship. 
Cybernetic ideas also found their way into the built environment with responsive architecture~\citep{Negroponte1975,ResponsiveArchitectures2006}, where sensors enable buildings to adapt to their environment and current conditions.

With the propagation of personal computers~\citep{Pagels1989}, the scientific study of complex systems~\citep{Bar-Yam1997,Mitchell:2009} continued the cybernetic tradition of studying phenomena in terms of their properties and functions. More recently, network science has provided tools for studying the components of complex systems (represented as nodes) and their interactions (represented as links)~\citep{NewmanEtAl2006,Newman:2010}.
This has allowed the application of concepts developed in different disciplines -- including cybernetics, complex systems, and network science -- to the understanding of urban systems~\citep{batty2005cities,Portugali2012,Batty2013,Bettencourt2013}. As technology has progressed, there have been several examples of the benefits of adaptivity in urban systems~\citep{Gershenson:2013}.

\begin{figure}[htbp]
\begin{center}
\includegraphics[width=0.45\textwidth]{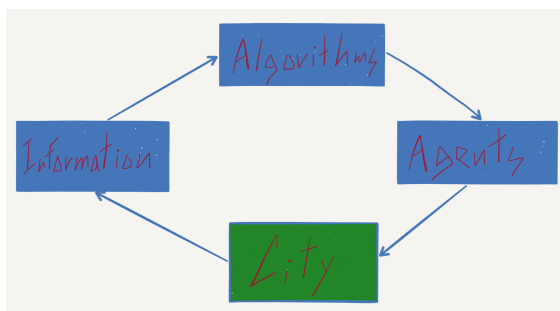}
\caption{Requirements of adaptive cities: an urban control loop.}
\label{fig:diagram}
\end{center}
\end{figure}

In this paper we sketch an urban theory that addresses the requirements to build \emph{adaptive cities}, their features, and their effects. We divide the requirements in three components: information, algorithms, and agents. These loosely correspond to the cybernetic sensors, control, and actuators, as illustrated by Figure~\ref{fig:diagram}. Traditionally, humans have fulfilled the roles of information, algorithms, and agents. However, advances in technology are assisting or replacing humans in different aspects of this ``urban control loop".

In the next sections, we detail information, algorithms, and agents, for then presenting the implications of building such adaptive cities.


\section{Information}

Information can be understood as anything that an agent can sense, perceive, or observe~\citep{Gershenson:2007,sloman2011s}. This is in accordance to Shannon's~\citeyearpar{Shannon1948} definition, in the sense that receiving information reduces uncertainty~\citep{ProkopenkoEtAl2007}. Any system requires information about the situation it is facing to make better-than-chance decisions. This is clear in animals, but applies to any system. Without relevant information, how could a system make the correct choice from a variety of potential decisions?

In urban systems, there are different sources of information which can be exploited for different purposes, such as measuring pollution or detecting traffic jams. ``Smart city" initiatives have integrated sensors pervasively~\citep{ETT:ETT2704},
 from parking spaces~\citep{sfPark2013} to trash bins
~\citep{6603732}. On the one hand, sensor and ICT costs are being reduced. On the other hand, the number of devices connected to the internet of things (IoT)~\citep{Sarma2000,Gershenfeld2004,Atzori2010IoT} is increasing. This creates the opportunity of obtaining ``big data" at a scale never before possible~\citep{Batty01112013}.

Nevertheless, most urban sensors and the information they generate have been the property of private companies. As an alternative, some cities have crowdsourced their information collection. For example, the City of Boston released the app Street Bump~\citep{carrera2013streetBump} to allow drivers to use their smartphones to automatically report potholes. This ``data donation" approach~\citep{AUFA2014} reduces the cost of sensor deployment, as most citizens carry potential sensors in their pockets~\citep{Ratti2006Mobile-Landscap,Gonzalez2008Understanding-i} and smartphones are becoming increasingly pervasive. Still, the massive adoption by citizens of the data donation approach is a major obstacle for cities obtaining relevant information ``for free". There has been a certain success of crowdsourced information~\citep{Lee2016} with platforms such as Ushahidi~\citep{okolloh2009ushahidi,Marsden201352} or Influenzanet~\citep{Paolotti2014}, but in these report-based initiatives information has to be entered by a human, thus limiting the amount and speed of information received. And simply data mining big data can be misleading~\citep{Lazer2014}.
Gamification has also been used to promote user participation, although with limited success~\citep{Odobasic2013}.

Another approach for obtaining information is from data citizens voluntarily publish on social networks such as Twitter~\citep{Bollen:2011,Dodds2011,Bertrand2013,Pina2016}. This research has focussed on detecting moods and emotions, but it has potential of expanding to gain insights into further aspects of urban and social life~\citep{Axhausen01122008,Cho:2011:FMU:2020408.2020579}.

A relevant aspect concerning urban information is that of privacy~\citep{Helbing2011,lane2014privacy,Enserink490}, as it has been shown that under certain conditions few data is required to uniquely identify citizens~\citep{Montjoye2013Unique-in-the-C,Montjoye:2015aa}. Information is required for adaptive cities, but there are potential risks if certain information becomes public. This creates a tension between efforts which strive for open access to information and individual privacy, which have led to proposals for data anonymization~\citep{Ghinita:2007:FDA:1325851.1325938,Montjoye2014} and self-regulatory information sharing~\citep{Pournaras2016}.

\section{Algorithms}

We are having more and more information at our disposal. But what to do with it?~\citep{Harford2014}. To put it in another way, what is the best way of extracting \emph{meaning} out of this information? One approach would be to use artificial intelligence to process information~\citep{Arel2010}. Still, many algorithms have been proposed and it seems that each is useful in a particular context. This is, there is no general detailed recipe for understanding all possible information that sensors could gather. Moreover, there are reasons to believe that there will never be one~\citep{wolpert95no,wolpert1997no}. 

There are deeper limitations that go beyond not having a general recipe for solving problems. Even if we had all relevant information in real time about an urban system, our predictability is limited by their complexity~\citep{GershensonHeylighen2005}: the components of urban systems are constantly interacting, and these interactions produce novel information which cannot be found in initial or boundary conditions~\citep{Gershenson:2011e}. Since our predictability is limited, optimization has to be complemented with adaptation~\citep{Gershenson2013Facing-Complexi}.

We might not have ready-made algorithms to solve every possible problem. But we can have methodologies which will assist their design. One of such methodologies is based on self-organization~\citep{GershensonDCSOS}. An example of self-organization can be seen with collective motion~\citep{Vicsek2012}: elements interact to generate a global pattern. This pattern is not determined by a single individual nor by an external source, but is the product of the interactions of the elements. The components generate the pattern, but the pattern also regulates the components. A system described as self-organizing is one in which elements interact in order to dynamically achieve a global function or behavior. Self-organization can help us build adaptive systems, as elements can self-organize when the conditions change, at the same timescale at which changes occur. Thus, with self-organizing algorithms we can face the unpredictability of urban systems inherent in their complexity~\citep{Ottino2004,Frei:2011,Yamu15102015}.

Examples of algorithms that use self-organization to regulate urban problems have already been proposed for coordinating traffic lights~\citep{Gershenson2005,Lammer:2008}, regulating public transport~\citep{Gershenson:2011a}, logistics~\citep{Helbing2006Self-organizati},  managing human organizations~\citep{GershensonSOBs}, and synchronization in power grids~\citep{Rohden2012}, among others. They are promising for building adaptive cities~\citep{Gershenson:2013,Yamu15102015,Rauws22072016}, as they use real-time information and self-organizing algorithms to achieve an efficiency close to optimal~\citep{Zubillaga2014Measuring-the-C,Viragh2014}, 
 or even supraoptimal~\citep{Gershenson:2011a,Tachet2016}, \emph{i.e.} the performance being better than what traditional theories predicted. 

Nevertheless, self-organization has also its limitations. If all control is based on local information, sometimes the performance will be worse than considering global information. This depends on the precise problem, and finding the right balance between local and global control is one of the challenges of adaptive systems~\citep{Forrest:2016:ACM:2975594.2964342} (see law of requisite variety below). Still, we know that once this balance is found, the system will adapt in the best possible way.

\section{Agents}

Information processed by algorithms has to impact urban systems through agents, \emph{i.e.} entities that \emph{act} on their environment. Citizens are certainly agents, as we constantly act on the urban fabric. However, we are embedded in sociotechnical systems which constrain and promote our actions and also can produce actions of their own~\citep{Vespignani2012,helbing2015thinking}. 

Sociotechnical systems are highly complex~\citep{Vespignani2009}, and this complexity limits their predictability. Not only interactions between components of urban systems generate novel information, but agents acting on cities will change the environment for the rest of the agents. Actions are clearly essential, but in many cases solutions can potentially generate novel problems. This is another reason for requiring adaptation in urban systems: if agents are changing urban problems while trying to solve them, our solutions must adapt to the changes that they themselves induce.

Technology is not only increasing the information we can collect, but also the agency of humans. We are coordinating actions in ways which were not possible only a few years ago~\citep{Marsden201352}. This capability has been useful for disaster response and has the potential of improving other urban systems, empowering citizens to act exploiting information and algorithms. 

Artificial agents in cities have been increasing their degree of autonomy for decision-making in recent decades. For example, traditional traffic lights have to be setup by humans. Semi-autonomous systems allow for the adjustment of phases of traffic lights, with the supervision and potential override by humans. Fully autonomous systems do not require human intervention to operate and adapt to changes in their environment~\citep{Gershenson2005}.

Perhaps the largest transformation we are witnessing in cities is related to the automation of vehicles. Their potential impact on urban mobility is manifold, as they promise to increase safety and efficiency while reducing emissions and congestion. Still, there are many open questions on the precise way in which autonomous vehicles will be introduced into cities. Will they be owned by individual? Shared by companies? Used as public transport by city governments? Probably all of them, but the most appropriate balance still has to be decided.

Autonomous vehicles are promising not only for the transportation of citizens, but also for logistic and delivery services. And they are not restricted to cars, as autonomous boats and drones will probably find their niche as well.

From this perspective of autonomous vehicles, how much of a city can be automatized? Could a city self-regulate most of its systems? To do so, information, algorithms, and agents must be integrated properly, as it will be discussed in the next section.

\section{Implications}

If we want to build an adaptive urban system, we must ask ourselves: how will we obtain relevant information? which algorithms will we use? which agents will act on the city? If we do not have a clear answer to one of these questions, it will be serendipitous if our system performs as desired. And we need not only to have proper information, algorithms, and agents. These also must be integrated properly, \emph{i.e.} the information acquired from the city has to be available to algorithms, algorithms should coordinate agents, and agents need to act on the city.

If we manage to develop information, algorithms, and agents, and integrate them to solve an urban problem, what would be the outcome? What is the benefit of having adaptive urban systems?

Imagine we had all relevant information about urban mobility in a whole city: where every citizen and vehicle is and where are they heading. Combining historic and current information, we could develop self-organizing algorithms that can find the best possible route for every citizen, for every vehicle. If these algorithms manage to act on all citizens and vehicles (easier with autonomy), then we could say that such a city would have \emph{optimal mobility}. This mobility would be optimal not because there would be no waiting times, but because there would be no better option given the current circumstances of demand and infrastructure. Such a system could also detect where new infrastructure would have the greatest impact for improving urban mobility, or where it might be most fragile. These suggestions could guide cities in building the most efficient and resilient transportation systems possible. Technically, it could be done already. In practice, we do not have access to all relevant information, and it is not obvious that we will ever have it. We can understand why such a system would produce optimal mobility because of Ashby's law of requisite variety~\citep{Ashby1956,BarYam2004,Gershenson2015Requisite-Varie}: in order to respond to a given variety of states in its environment, a control system must have at least the same variety. In other words, a system must be able to distinguish all different possible states which require a different action. Note that variety grows exponentially, so it is unfeasible to directly specify the requisite variety of the controller. This is precisely why adaptivity is necessary. Systems with a large variety will often be in states never before visited. An adaptive controller does not require all states to be predefined to react in an efficient way, this is why algorithms are used instead of functions.

Humans have a limited variety, and our control is also limited. Technology is allowing us to leverage some of that variety. In this way, we can expand our control capabilities, by exploiting technology to do some of the controlling. Moreover, technology also can be used for coordination. This can combine the variety of several humans or artificial systems to tackle systems with even more variety.

More variety implies more complexity. This precisely requires an integration of information, algorithms, and agents. If a system to regulate urban mobility has at least the same variety as the whole transportation system, \emph{i.e.} all possible combinations, then it will be able to respond to all possible situations. Thus, it will always be optimal, given the circumstances. The same reasoning applies to any urban system: if through the proper integration of information, algorithms, and agents we can have at least the same variety as the urban aspect the system is trying to control, then our adaptive urban system will be optimal, \emph{i.e.} performing in the best possible way for the given circumstances. If a controller does not have enough variety (less than the controlled), then we can distribute control using self-organization, increasing effectively the variety across several controllers. The precise scales at which control should be applied will depend on the variety of the controlled at different scales.

Even if we manage to achieve such optimality with adaptive urban systems, caution must be taken. If we are considering only certain variables for optimization, it does not imply that we are solving a problem completely. For example, even if we achieve maximum efficiency in urban mobility, such a system would not solve social issues which are partly an outcome of the processes which shape a city. Integrating a broader set of variables in the development of adaptive system requires the communication between all sectors of society. We are still in the exploratory process for finding efficient ways of achieving such communication~\citep{Zuckerman2014} and promoting social participation~\citep{Pickard509}. This would certainly be necessary if we pretend to achieve ``optimal" governance or sustainability~\citep{Trantopoulos2011}.

Cities have made efforts in recent years to increase their sustainability and resilience~\citep{Stumpp2013}. The discourse on resilient cities has focussed mainly on hazards~\citep{Godschalk2003} and climate change \citep{newman2009resilient,prasad2009climate}, which can also benefit from adaptation as described here~\citep{Pickett2004}.

\section{Conclusions}

Adaptive cities have the potential of increasing quality of life for citizens~\citep{Ratti2016}. But how equitable this increase of quality of life will be? Will all citizens benefit? At what cost? This is relevant, because even when cities accumulate most of the wealth of the planet, they are also the loci of greatest inequality. The answers to these questions will depend on how the adaptive urban technology is implemented, regulated, and managed in each city, and how this technology relates to citizens. This will require the effective interaction of governments, companies, academia, and society, as each sector may have different perceptions of the best way of managing cities.

For example, autonomous vehicles have a great potential to improve urban mobility. However, will this technology benefit few private companies, and/or the majority of citizens? The same technology can enslave or emancipate; the difference lies on how we use it. And the question is not so much who owns the technology. The key is how much can it \emph{interact}. For example, initially, the Internet infrastructure was mainly owned by academic and government institutions. Now mainly private companies own the infrastructure. Still, their business models allow the Internet to be an open system with standards where new technology and applications can thrive. Dedicated short-range communications (DSRC) have been designed specifically for automotive communication. Still, there are important differences across countries which limit compatibility as global standards are lacking. If this does not change, it will be difficult for vehicles to communicate among themselves and with infrastructure. Imagine that each website would require a different browser. If interactions are not possible, the potential of urban systems will be limited. For public transportation systems, the GTFS standard has been adopted by most cities, allowing information to be shared and exploited for novel applications~\citep{antrim2013many}. The same would occur with standards in other urban systems. The future of urban systems will not depend so much on who owns them, but on how openly can we interact with them.


\section*{Acknowledgements}
We should like to thank Anthony Vanky for useful suggestions. C.G. was supported by CONACYT projects 212802, 22134, 260021, and SNI membership 47907.

\bibliographystyle{cgg}
{\scriptsize \bibliography{carlos,refs}
}

\end{document}